\documentclass[aps,prl,twocolumn,groupedaddress,floatfix,superscriptaddress,nofootinbib]{revtex4-1}
\usepackage{empheq}

\usepackage{graphicx}
\usepackage[hidelinks]{hyperref}
\usepackage{color}
\usepackage{xcolor}
\usepackage[T1]{fontenc}
\usepackage[utf8]{inputenc}
\usepackage{amssymb}
\usepackage{amsmath}
\usepackage{amsthm}
\usepackage{bbm}
\usepackage{mathtools}
\usepackage{float}
\usepackage{empheq}
\usepackage{rotating}
\usepackage{soul}
\usepackage[hmargin=1.5cm,vmargin=1.8cm]{geometry}

\usepackage{pdfpages}
\makeatletter 
\AtBeginDocument{\let\LS@rot\@undefined}
\makeatother
\usepackage{pgffor} 

\def\tr{\operatorname{tr}}
\def\idty{\mathbb{I}} 

\def\bra#1{{\langle#1\vert}}
\def\ket #1{\vert#1\rangle}

\def\tr{\mathop{\rm tr}\nolimits}

\def\pg{P_g} 
\newcommand{\X}[1]{A_{#1}} 
\newcommand{\Y}[1]{B_{#1}} 
\newcommand{\pvm}[2]{P_{#1|#2}} 
\newcommand{\base}{e}

\newtheorem{theorem}{Theorem}

\usepackage{hyperref} 

\begin{document}

\title{Computing secure key rates for quantum cryptography with untrusted devices}

\author{Ernest Y.-Z. Tan}
\thanks{Corresponding authors; these authors contributed equally to this work.}
\affiliation{Institute for Theoretical Physics, ETH Z\"{u}rich, Switzerland}
\email{ernestt@ethz.ch}

\author{Ren\'{e} Schwonnek}
\thanks{Corresponding authors; these authors contributed equally to this work.}
\affiliation{Naturwissenschaftlich-Technische Fakultät, Universität Siegen, Germany}
\affiliation{Department of Electrical \& Computer Engineering, National University of Singapore, Singapore}
\email{r.schwonnek@gmail.com}

\author{Koon Tong Goh}
\affiliation{Department of Electrical \& Computer Engineering, National University of Singapore, Singapore}

\author{Ignatius William Primaatmaja}
\affiliation{Centre for Quantum Technologies, National University of Singapore, Singapore}

\author{Charles C.-W. Lim}
\email{elelimc@nus.edu.sg}
\affiliation{Department of Electrical \& Computer Engineering, National University of Singapore, Singapore}
\affiliation{Centre for Quantum Technologies, National University of Singapore, Singapore} 


\begin{abstract}
Device-independent quantum key distribution (DIQKD) provides the strongest form of secure key exchange, using only the input-output statistics of the devices to achieve information-theoretic security. Although the basic security principles of DIQKD are now well-understood, it remains a technical challenge to derive reliable and robust security bounds for advanced DIQKD protocols that go beyond the previous results based on violations of the CHSH inequality. In this work, we present a framework based on semi-definite programming that gives reliable lower bounds on the asymptotic secret key rate of any QKD protocol using untrusted devices. In particular, our method can in principle be utilized to find achievable secret key rates for any DIQKD protocol, based on the full input-output probability distribution or any choice of Bell inequality. Our method also extends to other DI cryptographic tasks.
\end{abstract}

\maketitle

\section*{Introduction}

Device-independent quantum key distribution (DIQKD) considers the problem of secure key exchange using devices which are untrusted or uncharacterized \cite{pironio,vidick2014,eatqkd}. In this setting, security is based entirely on the observation of non-local correlations, which are typically measured by a Bell inequality \cite{Bell1964,BrunnerNLRev}. In particular, if the correlations violate a Bell inequality, then we say that they are non-local. This is useful for secure key distribution, for it certifies that the key must come from measurements on an entangled state \cite{HorodeckiEntRev,curty2004entanglement,acin2005quantum}. While the basic principle behind the security of DIQKD is well understood from the monogamy property of non-local correlations \cite{barrett2006maximally}, an explicit security analysis is rather involved and tricky.  This is mainly because the dimension of the underlying shared quantum state is unknown and therefore the usual security proof techniques can not be applied.

\begin{figure}[t]
	\includegraphics[width=0.7\linewidth]{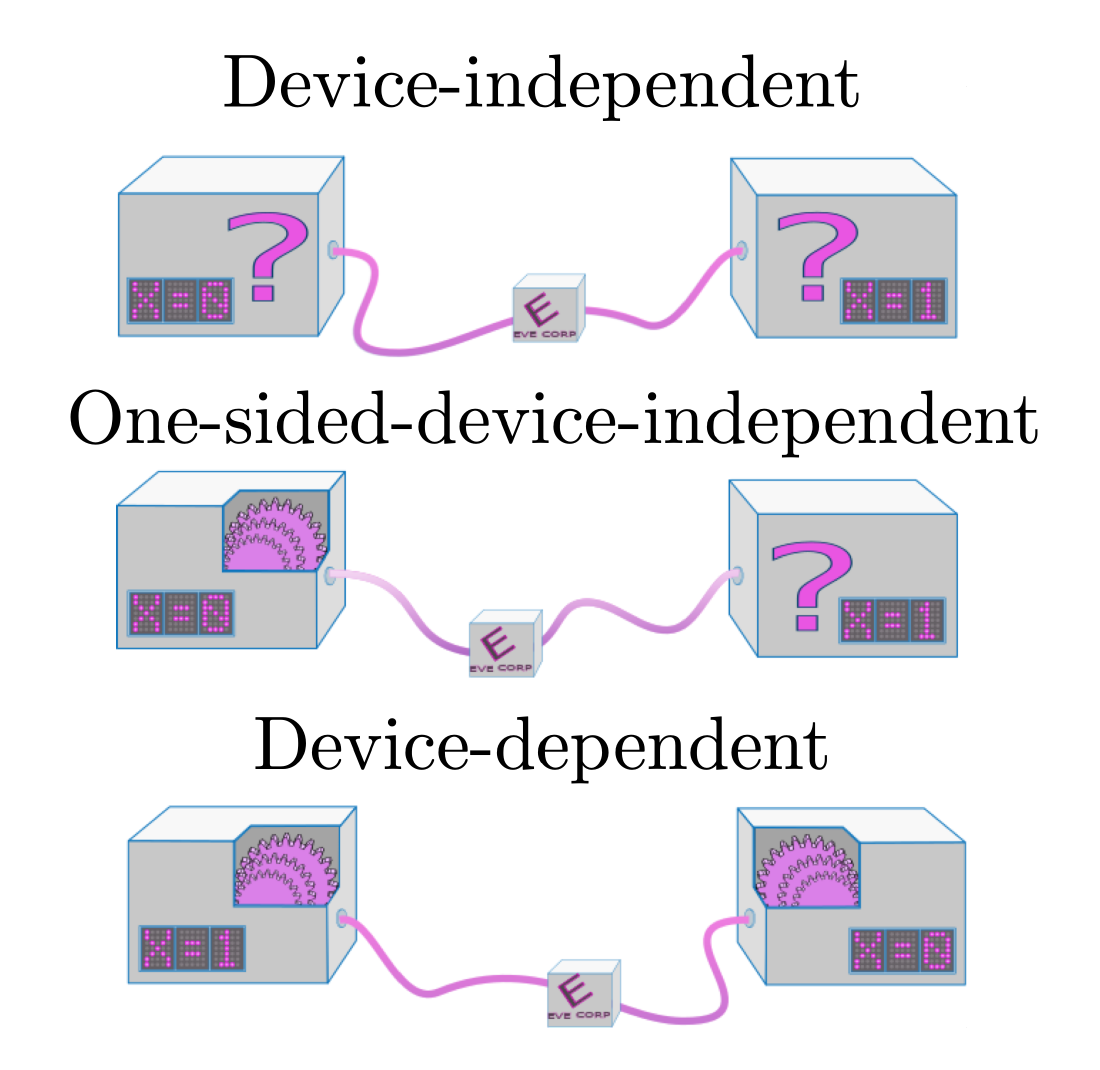}
	\caption{\textbf{Levels of device assumptions:} 
	Under device-dependent (DD) assumptions, all measurements and their underlying Hilbert spaces are characterized. Under fully device-independent (DI) assumptions, none of these are known, and we only assume the validity of quantum mechanics. {One-sided device-independent (1sDI)} assumptions lie between these two cases. 
	For the 1sDI setting, we consider the case where one party's measurements are fully characterized while the other's are unknown (e.g.~see Refs.~\cite{branciard2012one, tomamichel2013one}). 
	\label{fig:assumptions}}
\end{figure}

Recently, security proof techniques based on semi-definite programming (SDP) have been proposed for standard QKD \cite{coles16,winick,lin19,Wang2019,Primaatmaja2019}. In this so-called \emph{device-dependent} (DD) setting, the underlying QKD devices are assumed to be suitably characterized. Our main result extends this approach to a wider range of settings, adapting to different levels of device characterization (see Fig. \ref{fig:assumptions}). Previously, to prove the security of DIQKD, the existing approaches were to either employ a reduction to qubit-level systems \cite{pironio}, or to bound the adversary's guessing probability \cite{Masanes2011,bancal,nietosilleras14}. However, the former is restricted to protocols based on the CHSH inequality or similar Bell inequalities with binary inputs and outputs, while the latter only bounds the min-entropy, which typically leads to sub-optimal bounds on the von Neumann entropy (the relevant quantity for computing secret key rates against general attacks {in the asymptotic limit}~\cite{eatqkd}). The direct computation of DIQKD secret key rates is therefore an important task to address~\cite{pir2019advances}.

Here, we approach this problem with a universal computational toolbox that directly bounds the von Neumann entropy using the complete measurement statistics of a device-independent cryptographic protocol. Given this, our method not only applies to QKD, but also to some other DI cryptographic tasks such as randomness expansion~\cite{acin2016certified,pironio2010random,colbeckThesis,liu19,shalm19}. Importantly, this computational approach liberates the scope of device-independent cryptography to more complex scenarios, which could prove useful in analyzing the security of non-standard protocols which are known to be more robust against noise and loss~\cite{vertesi2010closing,froissart1981,sliwa2003,collins2004,gisin2009}, as well as multipartite protocols~\cite{ribeiro2018}.

The main mechanism of our toolbox is a technique for estimating the \emph{entropy production} of a quantum channel acting on an unknown state under algebraic constraints. {Entropy production}  \cite{davidlandauer,jarzynski,clausius,bekenstein} {is a fundamental concept traditionally used to study non-equilibrium thermodynamical processes, but here we show that it has an intrinsic connection to quantum cryptography as well.} The simplest way to understand entropy production is to view it as the amount of entropy introduced to a system after performing some action on it. For instance, in the case of projective measurement, the entropy production is the entropy difference between the post-measurement system and the initial system. 

Our toolbox bounds this entropy production via a (non-commutative) polynomial optimization over the measurement operators in the protocol. This can be evaluated using the SDPs in the the Navascu{\'e}s-Pironio-Ac{\'\i}n (NPA) hierarchy~\cite{npa}. In this context, switching from DI to {1sDI} or DD scenarios translates to adding more constraints on the SDPs and thus higher values for the final secret key rates. We present the key ideas used to derive this bound in the Methods section, and more specific details in Sec.~I--III of the Supplement.

(After release of this preprint, other approaches to solve the same optimization problem were separately developed in~\cite{brown21,brown21optimal}, with the technique in the latter yielding arbitrarily tight bounds in principle. We refer the interested reader to those works for comparisons and further details.)

\section{Results}
\subsection{Main theorem}

We focus mainly on describing our results for DIQKD, with results for other DI cryptographic tasks elaborated on in Sec.~I--IV of the Supplement.

\begin{figure}[h]
\includegraphics[width=0.9\linewidth]{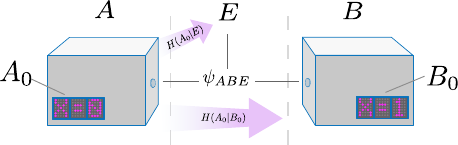}
\caption{\textbf{Basic situation:} By measuring her share of the joint state $\psi_{ABE}$ with measurement $\X 0$, Alice is (virtually) sending a raw key to Bob who (virtually) receives it by measuring $B_0$. Bob's uncertainty {about Alice's bit values} is quantified by the classical entropy $H(\X 0| B_0)$. We assume that Eve has access to all classical communication and her share of the joint quantum state, which gives her some partial information on $\X 0$ as well. This is quantified by the classical-quantum entropy $H(\X 0|E)$. \label{keysetting} 
}
\end{figure}

To assess the security performance of QKD, one can start by finding the asymptotic key rate under the assumption of independent and identically distributed (IID) states. In this setting, we consider protocols that are modelled as follows: in each round, Alice and Bob share a quantum state $\rho_{AB}$, and Eve's side-information $E$ is described by the purification $\psi_{ABE}$ of $\rho_{AB}$ (see Fig.~\ref{keysetting}). Qualitatively, this means Eve controls all systems that are not in the labs of Alice and Bob. In each round, Alice (resp.~Bob) performs one measurement from a set $\{\X{0},\X{1},\dots\,\X{\mathcal{X}-1}\}$ (resp.~$\{\Y{0},\Y{1},\dots,\Y{\mathcal{Y}-1}\}$) on 
their
local system. The raw key will be produced from the measurements $(\X{0},\Y{0})$. This model describes entanglement-based protocols, but can be easily converted to security proofs for prepare-and-measure protocols~\cite{shorpreskill,bbm92,Wang2019}. Here, we 
focus
on protocols that use one-way error correction. In this case, the asymptotic key rate $r_\infty$ is lower bounded by the Devetak-Winter formula~\cite{devetak}:
\begin{align}
r_\infty=\max\{H(\X{0}|E)-H(\X{0}|\Y{0}),0\}, \label{dwbound}
\end{align}
where $H$ is the von Neumann entropy {(which reduces to the Shannon entropy for $H(A_0|B_0)$)}. This can be intuitively interpreted as the difference between Eve's and Bob's uncertainty about Alice's measurement $\X{0}$. 

The $H(\X{0}|\Y{0})$ term in Eq.~\eqref{dwbound} can be computed based on the expected behaviour of the devices (see~\cite{eatqkd} for more details), so the main challenge here is to reliably bound $H(\X{0}|E)$ using the observed statistics. More specifically, suppose the protocol estimates parameters of the form $l_j = \sum_{abxy} c_{abxy}^{(j)} \mathrm{Pr}(ab|xy)$ for some coefficients $c_{abxy}^{(j)}$, where $\mathrm{Pr}(ab|xy)$ is the probability of obtaining outcomes $(a,b)$ from measurements $(\X{x},\Y{y})$ (e.g.~these parameters could be Bell inequalities in a DI scenario). Without loss of generality (see Sec.~V of the Supplement) we assume all measurements are projective by taking an appropriate Naimark dilation. For simplicity, we take the systems to be finite-dimensional; however, we do not impose any upper bound on the dimension. Let $\pvm{a}{x}$ denote the projector corresponding to an outcome $a$ of Alice's measurement $\X{x}$, and analogously, let $\pvm{b}{y}$ denote Bob's measurement projectors. Our task is to find lower bounds on 
\begin{equation}
\begin{gathered}
\inf H(\X{0}|E) \\
\text{s.th.~} \left\langle L_j \right\rangle_{\rho_{AB}} = l_j,
\label{eq:maintask}
\end{gathered}
\end{equation}
where $L_j = \sum_{abxy} c_{abxy}^{(j)} \pvm{a}{x} \otimes \pvm{b}{y}$, and the infimum takes place over $\psi_{ABE}$ and any uncharacterized measurements (which may be some or all of the measurements, for 1sDI or DI scenarios). 
This is a non-convex optimization (even after applying the approach from~\cite{coles16}), and the dimensions of any uncharacterized measurements could be arbitrarily large, hence there is no \textit{a priori} guarantee that any specific dimension suffices to find the minimum. 
Our central result is a method to tackle this task despite its challenges, 
which we achieve by proving the following theorem:

\begin{figure*}\centering
\includegraphics[width=0.9\textwidth]{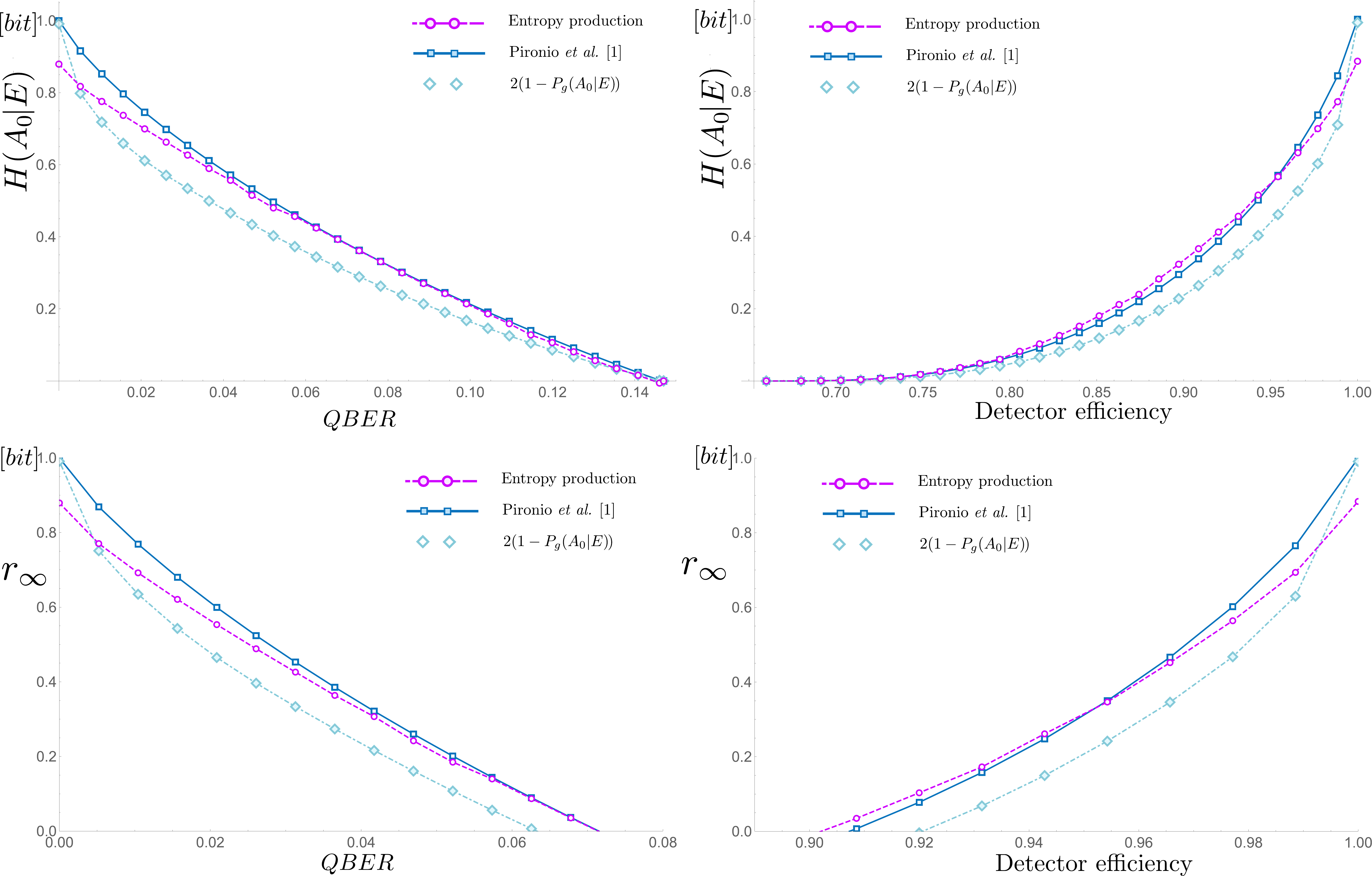}
\caption{\textbf{2-input 2-output DI protocols, $H(\X{0}|E)$ and $r_\infty$ (in base 2):} Lower bounds on entropy $H(\X{0}|E)$ and keyrate $r_\infty$, as a function of depolarizing noise (for the scenario studied in~\cite{pironio}) or detection efficiency (for the scenario studied in~\cite{eberhard93}, which optimizes the state and measurements to achieve maximal CHSH value). For the latter, $r_\infty$ was computed by optimizing the key-generating measurement $\Y{0}$ alone to minimize the value $H(\X{0}|\Y{0})$, without changing the state and other measurements from those in~\cite{eberhard93}. Also, to yield higher keyrates, the key-generating measurement $\Y{0}$ was preserved as a 3-outcome measurement (following~\cite{ma12}) rather than postprocessing it to 2 outcomes. It can be seen from the graph that our bounds are either close to or slightly better than the best previous result~\cite{pironio} for these scenarios, which was based on the CHSH value alone. For comparison, we also show the indirect bounds obtained by using the inequality $H(\X{0}|E) \geq 2(1-\pg(\X{0}|E))$ (in base 2).
}
\label{fig:DI}
\end{figure*}

\begin{theorem}
For a DI scenario as described, the minimum value of $H(\X{0}|E)$ (in base $\base$), subject to constraints $\langle L_j \rangle_{\rho_{AB}} = l_j$ with $L_j = \sum_{abxy} c_{abxy}^{(j)} \pvm{a}{x} \otimes \pvm{b}{y}$,
is lower-bounded by
\begin{align}
\sup_{\vec{\lambda}} \left(\sum_j \lambda_j l_j - \ln \left(\sup_{\substack{\rho_{AB},\pvm{a}{x},\pvm{b}{y} \\ \text{s.th.~} \langle L_j \rangle_{\rho_{AB}} = l_j}} \langle K \rangle_{\rho_{AB}}\right) \right),
\end{align}
where 
\begin{align}
K = T\left[
\int_\mathbb{R} dt \, \beta(t)
\left|\prod_{xy} \sum_{ab} \base^{\kappa_{abxy}} \pvm{a}{x} \otimes \pvm{b}{y} \right|^2
\right], \label{opKDI}
\end{align}
with $T[\sigma_{AB}]=\sum_a (\pvm{a}{0} \otimes \idty_B) \sigma_{AB} (\pvm{a}{0} \otimes \idty_B)$,
$\beta(t)=({\pi}/{2})(\cosh(\pi t)+1)^{-1}$, and $\kappa_{abxy} = (1+it) \sum_{j} \lambda_j c^{(j)}_{abxy} /2$. 
The integrals can be evaluated in closed form (we give the explicit expressions in Sec.~II of the Supplement).
\label{th_DIbound}
\end{theorem}

Importantly, Eq.~\eqref{opKDI} is a non-commutative polynomial in the measurement operators, and thus the task of bounding $\langle K\rangle_\rho$ 
can now be tackled using the well-established NPA hierarchy~\cite{npa}.
We can also study {1sDI} scenarios by imposing additional algebraic constraints corresponding to those satisfied by the characterized measurements. We highlight that since the optimization over $\vec{\lambda}$ is a supremum, any value of $\vec{\lambda}$ yields a \emph{secure} lower bound, without needing to identify the optimal $\vec{\lambda}$. 

To go beyond the asymptotic IID scenario,
one could apply the recently developed \emph{entropy accumulation theorem} \cite{eat,eatqkd}. This technique is applicable to DD, 1sDI and DI scenarios, and shows that the key rate against general attacks is still of a form essentially similar to Eq.~\eqref{dwbound}. It inherently accounts for finite-size and non-IID effects, and reduces the main challenge in a security proof to an IID problem, namely, 
finding lower bounds on the optimization problem in Eq.~\eqref{eq:maintask}
(see~\cite{eatqkd,eat,brown18} for more details).
{Specifically, our technique allows us to bound the \emph{min-tradeoff function} in the statement of the entropy accumulation theorem.} Hence our approach could also be used to compute finite key lengths against general attacks, by applying the entropy accumulation theorem.

\begin{figure*}
\centering
\includegraphics[width=0.99\textwidth]{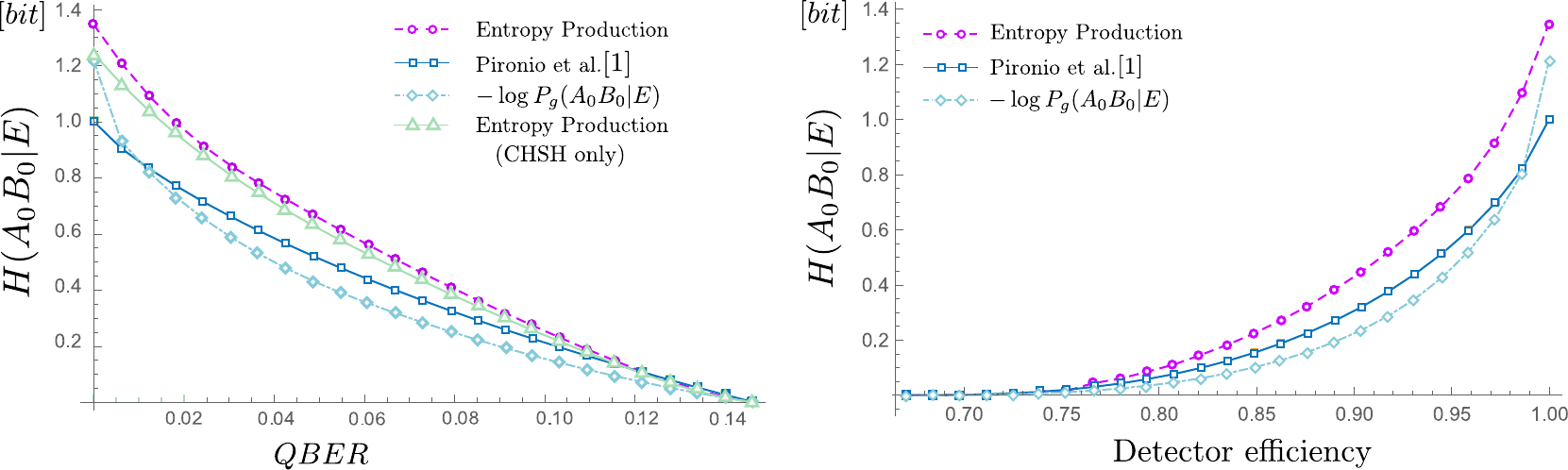}
\caption{\textbf{2-input 2-output DI protocols, $H(A_0 B_0|E)$ (in base 2):} Lower bounds on $H(A_0 B_0|E)$ as a function of depolarizing noise (for the scenario studied in~\cite{pironio}) or detection efficiency (for the scenario studied in~\cite{eberhard93}). Our approach outperforms both previous approaches, namely indirect bounds via the one-party entropy $H(A_0|E)$ (using the bound in~\cite{pironio}) or the guessing probability.
We also show a curve obtained when applying our method with only the CHSH value as the constraint, instead of the full output distribution.
}
\label{fig:DIRE}
\end{figure*}

\subsection{Computed keyrates}

We apply our method to two commonly studied DI scenarios, in which Alice and Bob each perform parameter estimation on 2 binary-outcome measurements. (For QKD purposes, Bob will need to perform a third measurement for key generation, corresponding to $\Y{0}$ in Eq.~\eqref{dwbound}, but we do not use this when bounding $H(\X{0}|E)$.) Our results are shown in Fig.~\ref{fig:DI}. Results in some other scenarios, including distributions optimized for tilted CHSH inequalities~\cite{acin12}, are presented in Sec.~IV of the Supplement.
The first scenario is parametrized by a depolarizing-noise value $q\in[0,1/2]$, and corresponds to performing the ideal CHSH measurements {(i.e. $A_0 = Z$, $A_1 = X$, $B_0 = (Z+X)/\sqrt{2}$) and $B_1 = (Z - X)/\sqrt{2}$)} on the Werner state $(1-2q)\ket{\Phi^+}\bra{\Phi^+} + (q/2)\idty$, where $\ket{\Phi^+}$ is the Bell state $(\ket{00} + \ket{11})/\sqrt{2}$ {and $Z$ and $X$ are Pauli operators}. The second scenario is a limited-detection-efficiency model parametrized by $\eta \in [0,1]$, where for every measurement the outcome $1$ is flipped to $0$ with probability $1-\eta$. This is a simplistic model for a photonic setup where all non-detection events are mapped to the outcome $0$~\cite{eberhard93}. For this scenario, we use different states and measurements for each value of $\eta$, as follows: to compute the $H(\X{0}|E)$ bound, we first optimize the state and parameter-estimation measurements to maximize the CHSH value the same way as in~\cite{eberhard93}; then to compute the $r_\infty$ curves, we optimized the key-generating measurement $\Y{0}$ on its own without changing the state or other measurements. In principle, this yields parameter choices that may be suboptimal for maximizing $H(\X{0}|E)$ or $r_\infty$, since maximizing either of these quantities is not necessarily equivalent to maximizing CHSH value (this was later confirmed in~\cite{brown21,brown21optimal}, which aimed to optimize the rates directly). However, our method is too computationally intensive to attempt to maximize $H(\X{0}|E)$ or $r_\infty$ directly, so we use the CHSH value as an indirect proxy (since it can be optimized independently of our bounds). 

The previous best bound on $H(\X{0}|E)$ in these scenarios (see Sec.~IV of the Supplement for known results in other cases) was that derived in Ref.~\cite{pironio}, which uses only the CHSH value instead of the full probability distribution. To make use of the latter, the only preceding approach was to first bound the guessing probability $\pg(\X{0}|E)$ and then apply the inequality $H(\X{0}|E) \geq -\ln\pg(\X{0}|E)$~\cite{Masanes2011, bancal,nietosilleras14} (all entropies are in base $\base$ unless otherwise specified). We note that if the marginal distribution of $\X{0}$ is uniform and binary-valued, then in fact the tighter inequality~\cite{briet09} $H(\X{0}|E) \geq (2\ln2)(1-\pg(\X{0}|E))$ holds, and we plot this bound in Fig.~\ref{fig:DI}. (See Sec.~IV of the Supplement for details on how it applies in the limited-detection-efficiency model.) However, approaches based on guessing probability do not outperform the bound in~\cite{pironio} for the two scenarios considered here.

Our method uses the full input-output distribution to bound $H(\X{0}|E)$ directly. As shown in Fig.~\ref{fig:DI}, it gives results that are close to or slightly outperform the bound from Ref.~\cite{pironio}. Roughly speaking, our approach tends to perform well for moderate noise values, which is useful since many Bell-test implementations are currently in such noise regimes~\cite{Hensen2015,Giustina2015,Shalm2015,Rosenfeld2017,Murta2018}. 
Our results prove that for the limited-detection-efficiency scenario, better bounds on $H(\X{0}|E)$ can be obtained by considering the full distribution rather than just the CHSH value (since the CHSH-based bound~\cite{pironio} is tight). This suggests it may not be optimal to simply choose experimental parameters that maximize the CHSH value---maximizing a different Bell value may allow our method to yield a further improvement over the results in Fig.~\ref{fig:DI}.

\begin{figure}[b]
\includegraphics[width=0.94\linewidth]{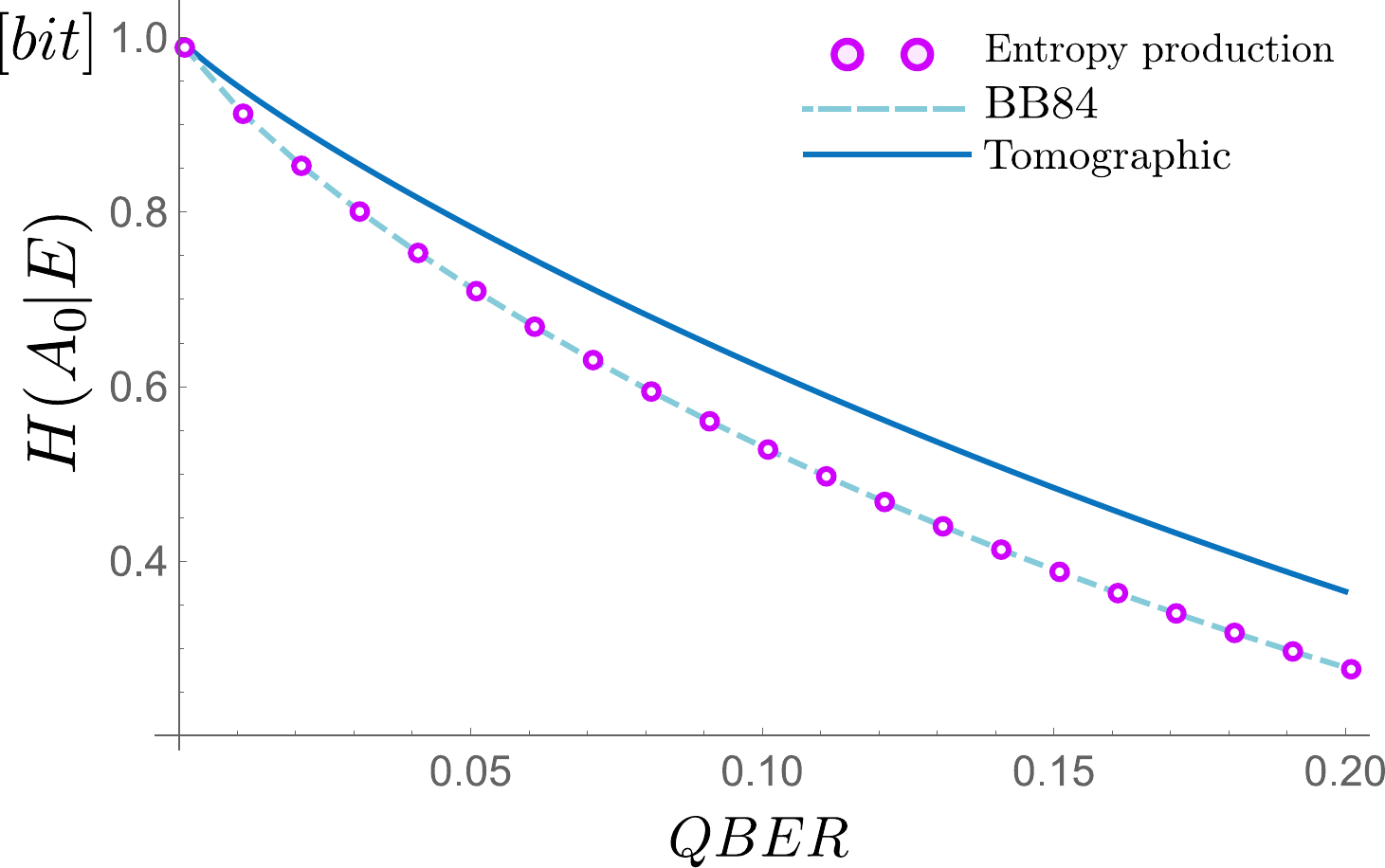}
\caption{\textbf{1sDI six-state protocol:}
Lower bounds on $H(\X{0}|E)$ for a {1sDI} version of the six-state protocol~\cite{ss}. Interestingly, the bound we obtain from our method coincides with that for the BB84 protocol. For reference, we also show the bound that could be obtained from a tomographically complete characterisation of the state, such as via the measurements in the standard (device-dependent) six-state protocol.}
\label{fig:sixstate}
\end{figure}

With minor modifications (see Sec.~I of the Supplement) our method can also bound the ``two-party entropy'' $H(\X{0}\Y{0}|E)$, which is relevant for DI randomness expansion~\cite{acin2016certified,pironio2010random,colbeckThesis,liu19,shalm19}. The previous approaches for this were similar to those for $H(A_0|E)$: firstly, simply noting that $H(A_0 B_0|E) \geq H(A_0|E)$ and then applying the bound from~\cite{pironio}; secondly, bounding it via $H(A_0 B_0|E) \geq -\ln \pg(A_0 B_0|E)$. These approaches are suboptimal for similar reasons as before, though here the former is further limited by the fact that it ignores the register $B_0$. As shown in Fig.~\ref{fig:DIRE}, our method clearly outperforms both of these approaches, which could improve the key rates for DI randomness expansion.

We also analyze a 1sDI version of the six-state protocol~\cite{ss}, where Bob's measurement device is uncharacterized. As mentioned earlier, the characterization of Alice's device translates to algebraic relations between the operators $\pvm{a}{x}$, which we impose as additional constraints on top of the NPA hierarchy. We see that in Fig.~\ref{fig:sixstate}, the resulting bound coincides with the bound for the BB84 protocol. This supports a conjecture~\cite{kt} that when Bob's measurements are uncharacterized, performing three measurements does not offer any advantage over performing only two measurements.

\section{Discussion}
Here, we have developed a universal toolbox to obtain reliable secret key rates for QKD with untrusted devices. The main advantage of our method is that it can be applied to any DIQKD protocol, not only those based on specialized Bell inequalities. The only previous known approach that could be applied to DIQKD with such generality is that based on bounding the guessing probability~\cite{Masanes2011, bancal,nietosilleras14}, which is generally not optimal. Our method outperforms all earlier results in some cases, as shown in Figs.~\ref{fig:DI}~and~\ref{fig:DIRE}. Importantly, it seems to give good bounds in regimes with substantial noise, which are likely to be experimentally relevant. 

Currently, our method scales rapidly in computational difficulty as the number of inputs or outputs for the protocol increases---the polynomial in Eq.~\eqref{opKDI} is generally of high order, hence a high NPA hierarchy level~\cite{npa} is needed to bound $\langle K\rangle_\rho$. Because of this, we currently do not have good bounds for DI scenarios with large numbers of inputs or outputs (though we find suboptimal bounds for some such cases; see Sec.~IV of the Supplement). 
An important goal now would be to find ways to improve the tractability of our approach, perhaps by following reductions along the lines of those described in Ref.~\cite{tavakoli19}. This would enable the computation of key rates for DIQKD protocols (or other DI protocols) with more measurement settings and/or outcomes.

With our toolbox in hand, one can now explore DI protocols based on maximizing a variety of Bell expressions (or maximizing the key rate directly) instead of being restricted to CHSH. While the scaling issues currently impose some limitations, we observe that there remains substantial unexplored territory even within 2-input 2-output DI protocols. For instance, the tilted CHSH inequalities~\cite{acin12} can certify higher two-party entropies than CHSH in the absence of noise, but the previous bounds were based on min-entropy and not very noise-robust. Using our approach to improve these bounds (see Sec.~IV of the Supplement) would be relevant for experimental implementations of DI protocols such as randomness expansion~\cite{liu19,shalm19}. 

\section{Methods}
\subsection{Bounding the von Neumann entropy}

\begin{figure}[h!]
	\includegraphics[width=0.8\linewidth]{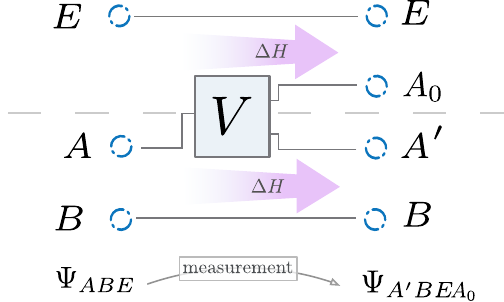}
	\caption{\textbf{Connection to entropy production:} The key-generating measurement is regarded as an isometry to a larger Hilbert space, by expanding the classical memory $A_0$ with an ancilla $A'$. From this perspective the initial and final states are pure, and thus the entropy change $\Delta H$ on the memory-Eve subsystem equals the entropy change on the Alice-Bob subsystem. \label{entropyflow}}
\end{figure}

The advantage of quantum over classical cryptography stems from the fact that for the former, it is possible to bound Eve's knowledge using only Alice's and Bob's systems (essentially, using the monogamy property of entanglement). To make this precise for $H(\X{0}|E)$, one can regard the key-generating measurement as a quantum-to-classical channel that maps Alice's (quantum) system $A$ to a memory register $\X 0$ which stores the (classical) measurement outcomes. By Stinepring's theorem \cite{stinespring1955positive}, this channel can be described via an isometry $V$ to an extended system $\X{0} A'$. This isometry maps the pure initial state $\Psi_{ABE}$ to a pure final state $\Psi_{A'BE\X{0}}$ (see Fig.~\ref{entropyflow}).

Since the entropies of the marginal states of a pure bipartite state are equal, this gives
\begin{align}
H(\X 0|E) &=H(\X{0} E)-H(E)\nonumber\\
&= H(A'B)-H(AB) \nonumber\\
&= H(T[\rho_{AB}])-H(\rho_{AB}) \eqqcolon \Delta H, \label{eq:condtoprod}
\end{align}
where $T[\rho_{AB}] = \tr_{\X{0}}((V \otimes \idty_B) \rho_{AB} (V \otimes \idty_B)^\dagger)$. (We remark that this approach was used in Ref.~\cite{coles11}.) The last line can be interpreted as entropy production, $\Delta H$, resulting from the transformation $AB\to A'B$. Since it only depends on the reduced states of Alice and Bob, they can be used to bound Eve's knowledge using only their own systems. For projective measurements, $V$ can be chosen~\cite{coles11} such that $T$ is the pinching channel
\begin{align}
    T[\rho_{AB}]=\sum_a (\pvm{a}{0} \otimes \idty_B) \rho_{AB} (\pvm{a}{0} \otimes \idty_B).
\end{align}

Besides its application to QKD, the amount of entropy that is produced or consumed by a quantum operation $T$ is one of the central quantities of a physical system.
However, computing this entropy quantity is technically challenging, since the entropy of a quantum state is not directly accessible. Instead, the quantities that are directly accessible are typically the expectation values of certain observables, i.e.~expressions of the form $\langle L_j\rangle_{\rho}=\tr(\rho L_j)$ for operators $L_j$ (which in QKD scenarios have the form described earlier). Following this perspective, we have to study the following problem: find bounds on $\Delta H$ that hold for all states consistent with the observed constraints $\langle L_j\rangle_\rho = l_j$. For QKD, these bounds have to be lower bounds, since we consider the ``worst-case scenario'' for the honest parties. 

To solve this problem, we propose the following \emph{ansatz}: for coefficients $\lambda_j \in \mathbb{R}$, we define $L=\sum_j \lambda_j L_j$ and aim to find an operator $K$ such that
\begin{align}
H\left(T[\rho]\right)-H(\rho)\geq\langle L\rangle_\rho-\ln\langle K \rangle_\rho,
\label{entrobound}
\end{align}
holds for all states. 
To find such a $K$, we note that Jensen's operator inequality and the Gibbs variational principle imply (see Sec.~III of the Supplement for details)
\begin{align}
H\left(T[\rho]\right)-H(\rho) &\geq -\left\langle\ln T^*T[\rho]\right\rangle_\rho - H(\rho)\nonumber\\
&\geq \langle L\rangle_\rho-\ln\tr\left(\base^{\ln(T^*T[\rho]) + L}\right),\label{gibbs-jensen}
\end{align}
where $T^*$ is the adjoint channel of $T$. Applying a recently discovered generalisation of the Golden--Thompson inequality~\cite{sutter}, it follows that for any self-adjoint {$\tilde{L}_k$} such that {$L=\sum_k \tilde{L}_k$}, we can choose
\begin{align}
    K = T^*T\left[ 
    \int_\mathbb{R} dt \, \beta(t)
    \Bigg|\prod_{k} \base^{\frac{1+it}{2} \tilde{L}_k} \Bigg|^2 
    \right], \label{opKgeneral}
\end{align}
where {$\beta(t)=({\pi}/{2})(\cosh(\pi t)+1)^{-1}$}. Thus, this yields a family of lower bounds on $H\left(T[\rho]\right)-H(\rho)$, characterized by $\lambda_j$ and {$\tilde{L}_k$}.

Our task is now reduced to finding upper bounds on $\langle K\rangle_\rho$. 
If the explicit matrix representation of $K$ is known, such as in a DD scenario, this is an SDP in a standard form and can be solved directly (see e.g.~\cite{coles16}).
However, 1sDI and DI scenarios appear much more challenging, because 
one does not have an explicit form for $K$. This reveals the key breakthrough allowed by our approach: a careful choice of $\tilde{L}_{xy}$ lets us bound $\langle K\rangle_\rho$ \emph{without} an explicit matrix representation. Specifically, by setting 
\begin{align}
\tilde{L}_{xy} = \sum_{abj} \lambda_j c^{(j)}_{abxy} \pvm{a}{x} \otimes \pvm{b}{y},
\end{align}
we obtain (see Sec.~III of the Supplement) Theorem~\ref{th_DIbound} as stated above. For the DI scenario, the channel $T$ is self-adjoint and idempotent, so $T^*T = T$. With this choice of $\tilde{L}_{xy}$, we achieved the critical goal of reducing $\langle K\rangle_\rho$ to a form that can be bounded using the NPA hierarchy.

\section{Data availability}
The data produced in this work is available from the corresponding authors upon reasonable request.

\section{Code availability}
The code used in this work is available from the corresponding authors upon reasonable request.

\section{Acknowledgements}
\begin{acknowledgments}
We thank Otfried Gühne, Miriam Huang, Mathias Kleinmann, Jie Lin, Norbert L\"{u}tkenhaus, Tobias Osborne, Gláucia Murta, Miguel Navascu{\'e}s, Renato Renner, Valerio Scarani, Marco Tomamichel,  Reinhard F. Werner, and Ramona Wolf for useful discussions. E.~Y.-Z.~T.~was funded by the Swiss National Science Foundation via the National Center for Competence in Research for Quantum Science and Technology (QSIT), and by the Air Force Office of Scientific Research (AFOSR) via grant FA9550-19-1-0202.
C.~C.-W.~L.~acknowledges support by the National Research Foundation (NRF) Singapore, under its NRF Fellowship programme (NRFF11-2019-0001) and Quantum Engineering Programme (QEP-P2), and the Asian Office of Aerospace Research and Development (FA2386-18-1-4033).

Computations were performed using the MATLAB package YALMIP~\cite{yalmip} with solver MOSEK~\cite{mosek}. Some of the calculations reported here were performed using the Euler cluster at ETH Z\"{u}rich.

\end{acknowledgments}

\section{Competing interests}
The authors declare no competing interests.

\section{Author contributions}
E.~Y.-Z.~T.~and R.~S.~are co-first authors on this work. Both contributed to the derivation of the main theorem, with inputs from all other authors. E.~Y.-Z.~T.~implemented the computations for DI scenarios, and R.~S., K.~T.~G., I.~W.~P., and C.~C.-W.~L.~studied 1sDI scenarios. C.~C.-W.~L.~and E.~Y.-Z.~T.~proposed the project and structured the overall concept. All authors contributed to writing the manuscript.

\section{Publication}

This version of the article has been accepted for publication, after peer review, but is not the Version of Record and does not reflect post-acceptance improvements, or any corrections. The Version of Record is available online at: \href{https://doi.org/10.1038/s41534-021-00494-z}{https://doi.org/10.1038/s41534-021-00494-z}

\foreach \n in {1,...,13} {\clearpage \includepdf[pages={\n}]{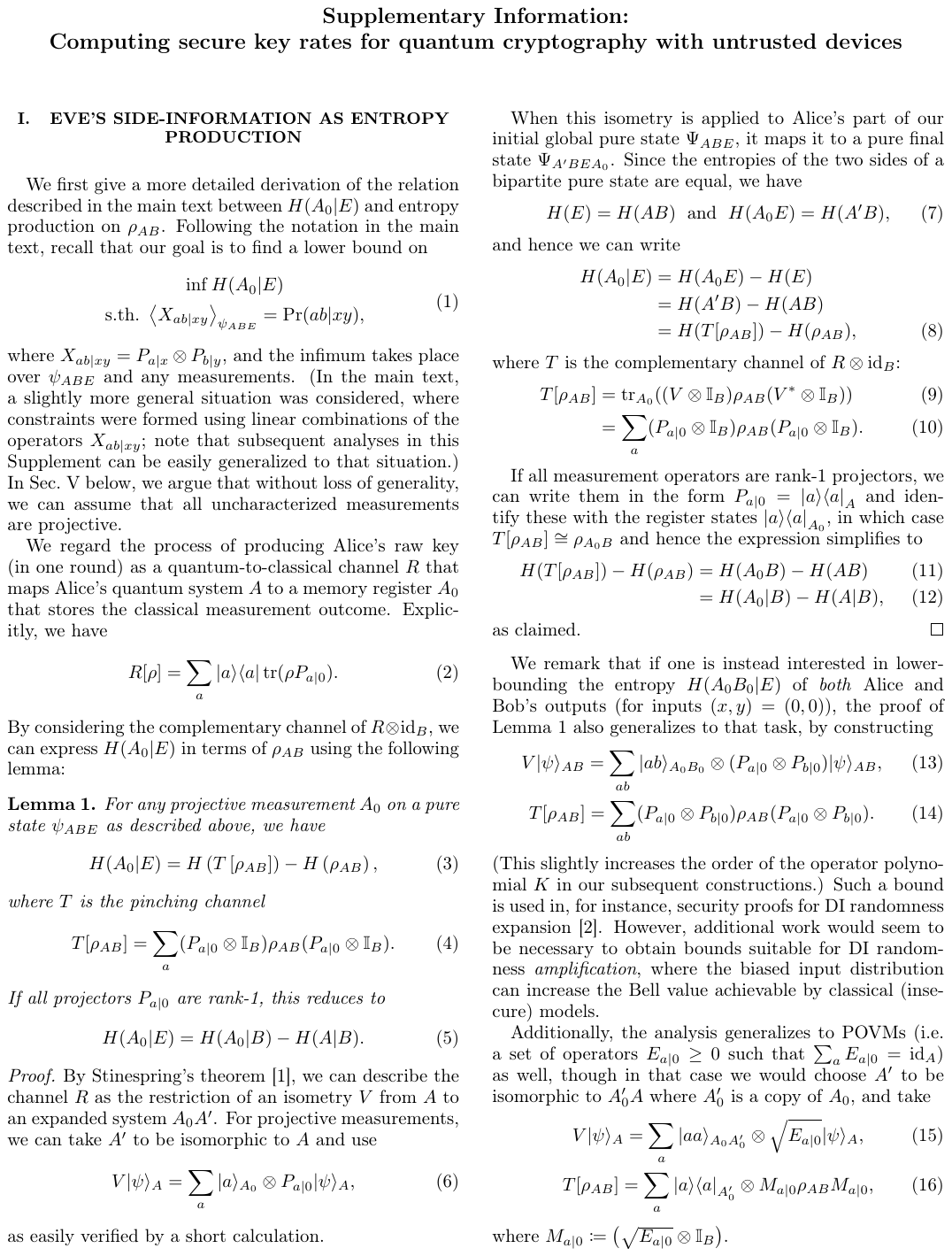}}

\end{document}